# Real-Time Recognition of Vortex Beams Modes Through Random Diffusive at the Speed of Light

Tong Fu[1], Gang Luo[1], Jia Cheng Li[1], Yuan Chao Geng[1*], Xiao Dong Yuan[1†]
[1] Laser Fusion Research Center, China Academy of Engineering, Mian Yang, 621900

## ABSTRACT

Optical vortex beam with orbital angular momentum (OAM) has great potential to increase the capacity of optical communication and information processing in classical and quantum regimes. Nevertheless, important challenges that influence the optical data transmission in free space is the existence of diffusers along the optical path, which causes inevitable information loss during the wave propagation. Numerous algorithms have been proposed successively for identifying the modes of vortex beams propagating through scattering media. However, these methods all require completion on a computer, which is energy-intensive and energy consuming. Here, we propose an all-optical regime for identifying the modes of vortex light fields propagating through scattering media. After training by deep learning, our model can recognize the mode of vortex beam through unknown phase diffusers, demonstrating generalization to new random diffusers that have never been encountered before. Once physically deployed, the entire setup will rapidly identify the modes of vortex light propagating through scattering media at the speed of light, and the entire inference process will consume zero energy except for illumination source. Our research represents a significant step towards highly accurate recognition of vortex light modes propagating through complex scattering media, providing significant guidance for the application of optical communication in complex environments.
**Keywords:** neuron network, vortex beams, orbital angular momentum, optical communication

## 1. INTRODUCTION

Vortex beams, possessing a helical phase front described by the phase factor $\exp(il\vartheta)$, have been demonstrated to have helical phase fronts and carry orbital angular momentum (OAM) $l\hbar$ per photon. Where $l$ is the topological charge representing OAM mode, $\hbar$ is the Plank constant $h$ divided by $2\pi$ [1], $\vartheta$ is the azimuth angle which determines the direction of angular momentum within the light beam. Owing to their distinctive optical properties, they have been extensively applied in optical communication [2-4], quantum information processing [5, 6] and quantum entanglement [7-9], optical trapping [10], etc. In optical communication, specifically, the orthogonal orbital angular momentum (OAM) modes provide infinite channels for signal multiplexing, thereby substantially enhancing communication capacity density [11-16]. In a notable milestone, Graham et al. were the pioneers in demonstrating the viability of leveraging orbital angular momentum states for optical information transmission in free space [17]. Krenn et al. made a groundbreaking stride by experimentally showcasing the identification of 16 distinct OAM modes traveling across a 3 km intra-city optical link [18], this methodology involved the utilization of an artificial neural network to discern the characteristic mode patterns exhibited on a screen at the receiver. Subsequently, various techniques, including holographic gratings, interferometers, and others, have been designed for detecting the orbital angular momentum modes of vortex light [19-24]. All the methods pave the way for enhancing the capacity of future optical communication and information processing.

Nevertheless, identifying vortex beam modes propagating through complex transmission media remains a significant challenge in the area of optical communication [25-27]. One of the obstacles that affecting the high-precision transmission of vortex beams (VBs) information in free space is the occurrence of random diffusive media along the optical pathway [28-29]. This inevitably results in optical information loss due to distortion in the optical wavefront during propagation. Several methods have been developed to mitigate the impact of such random diffusers, including phase recovery carried out using reconstruction algorithms [30, 31], However, prior information and digital computers are required to recover the phase distorted by diffusers, introducing additional complexity to the phase recovery process. Another method involves using adaptive optical to correct the phase distortion caused by diffusers along the optical path [32]. However, the utilization of spatial light modulators and feedback algorithms in adaptive optics increase both the complexity and cost of such systems. In recent years, propelled by the widespread application of deep learning, researchers have increasingly utilized deep learning algorithms in optical information transmission through scattering media. [33-36]. This has increased the recognition speed of orbital angular momentum and enhanced robustness against distortions caused by random diffusers. However, all algorithm implementations and optical information processing require encoding and extracting feature information on digital computers, which energy-intensive and time consuming. Recently, diffractive deep neural networks (D2NNs) have garnered considerable attention owing to their distinct advantages in speed, parallelism, and energy efficiency compared to traditional neural networks operating with electrons computers [37-40]. With powerful optimization functions and mapping capabilities, D2NN can precisely fine-tune the modulation parameters of each grid on diffractive layers, that influencing the wavefront of secondary wave sources to execute specific optical transformation, and widely used in pattern recognition, holographic imaging, optical decoding, etc. However, most of research have been focused on processing the intensity distribution of input optical field, the impact of diffusers along the optical path on the recognition of phase information has rarely been taken into account.

---

* gengyuanchao@caep.cn
† yxd66my@163.com

In this work, we propose an all-optical method for recognizing orbital angular momentum (OAM) modes, which does not rely on any electronic computers and algorithm during task execution. This method is capable of recognizing the mode of distorted vortex beam caused by unknown, randomly diffusers at speed of light, as shown in Fig1. Different from previous studies that utilized electronic computers to recover the distorted phase of vortex light [41], here we employ a deep neural network to train a series of diffractive modulation layers to optically recognize the patterns with different orbital angular momenta through random diffusers with the same correlation. Each diffractive layers that is trained has tens of thousands of neurons, and the phase and amplitude values of these neurons are adjustable parameters that is tuned during the training phase by using error back propagation algorithm. During training phase, VBs with $l \in [-5,5]$ are fed into network, A series of randomly generated phase diffusers are incorporated into the network to simulate scattering media along the optical paths, enhancing the generalization capability of optical neural networks. After a design process grounded in deep learning, the layout of passive diffractive modulation layers are constructed. In these layers, the transmission or reflection coefficients of each neuron in all layers remain fixed. These layers are then fabricated to form a physical diffractive network positioned between randomly generated phase diffusers and the output plane. As vortex beams (VBs) carrying orbital angular momentum (OAM) information propagate through these previously unseen diffusers, the distorted wavefront is collected and modulated by a trained diffraction network. This enables all-optical inference of the OAM mode at its output field of view, without any need for analog-to-digital conversion or digital computing.

This article is structured as follows: Section 2 elaborates on the diffraction principles of optical neural networks in details, Section 3 presents our simulation results and discusses them, and finally, the research conclusions are provided. Although the proposed all-optical diffractive network-based OAM mode recognition framework considers only a thin layer of random diffusers, it has the potential to be extended for recognition through volumetric diffusers. Such an extension could have transformative impacts in various fields, where OAM mode reorganization through diffusive media is critical. Although this study is confined to optical network design and theoretical simulations, but we have developed a method capable of identifying vortex light in complex transmission environments, provided a reliable theoretical basis for further experimental implementation. We believe that our regime of recognition of OAM mode using diffractive neuron network will provide significant guidance for the advancement of optical communication.

## 2. MODEL AND METHOD

### 2.1 The forward propagation models

The framework of four layers diffractive networks for OAM mode recognition is shown in Fig.1. In terms of network layout, randomly phase diffusers with same correlation length are individually placed away from the input plane, the phase diffusers are modeled as the phase only mask that distorting the wavefront of VBs, the distorted vortex light is then transferred into the trained diffraction network for phase modulation. The neurons on each diffractive layer are interconnected through optical transmission. During the inference process, the distorted light field are modulation by the trained layered structure, ultimately converging at specific positions on the detection plane to characterize the patterns of the vortex light. The entire process is completed at the speed of light. Note that the number of hidden layers in Fig.1 are only illustration, which can be any integer greater than or equal to 1. By tuning the structural parameters of each neuron during the training process, the phase values in each neuron are fixed.

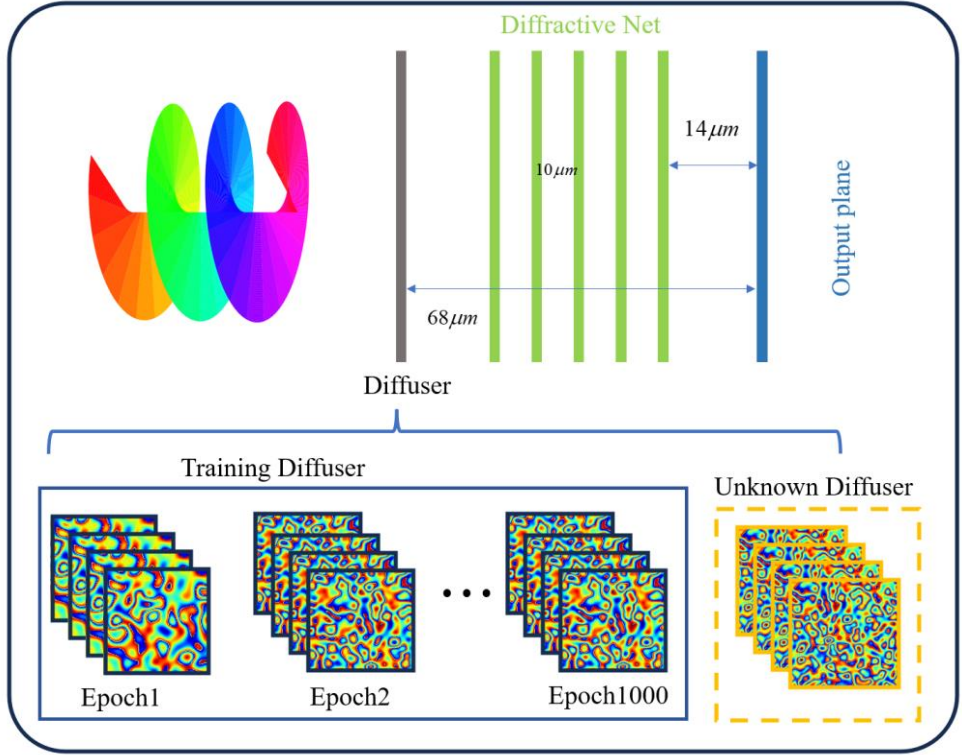

Figure 1. Schematic of all-optical deep neuron network for recognize the OAM mode under the condition of scattering media. The distorted phase distribution of VBs behind a random phase diffusers are converted into a normalized intensity distribution. The bottom row shows the random phase distribution with the same correlation length $L \sim 7\lambda$ in traing and inference processes respectively.

The basic physics of the modulate layers is discussed the following, according to Huygens-Fresnel principle, each neuron on the modulate layers acting as a secondary source of a wave, the shape of the new wavefront at the subsequent moment is determined by the envelope of the secondary spherical wave. Therefore, each unit on the diffractive layers can be considered as a neuron fully connected the proceeding neurons and the following neurons. The optical interconnection between layers can be modeled as Rayleigh–Sommerfeld diffraction integral:

$$h_l^i(x, y, z) = \frac{z - z_i}{r^2} \left( \frac{1}{2\pi r} + \frac{1}{j\lambda} \right) \exp\left( \frac{i 2\pi r}{\lambda} \right) \qquad (1)$$

Where $l$ represent $l-th$ layer of the network, $i$ represents $i-th$ neuron located at $(x_i, y_i, z_i)$ of layer $l$, and for $l = 1$ denotes the input layers, and $r = \sqrt{x^2 + y^2 + z^2}$ denotes the distance between neurons. $\lambda$ is the illumination wavelength, the amplitude and relative phase of this secondary wave are dictated by the product of the input wave to the neuron and its transmission coefficient. Considering VBs $f(x, y, z = 0)$ was used as input field at input plane, propagate along the optical path and distorted by random diffusers:

$$f(x, y, z = z_0) = t_f(x, y) \cdot [f(x, y, z = 0) * h_l^i(x, y, z)] \qquad (2)$$

Where $z = z_0$ is the axial distance between input plane and the position of diffusers, and $t_f(x, y)$ denotes random phase diffusers that distort the wavefront of optical field, this distorted optical field is used as input field of the diffractive network. And the diffractive is modeled as a series of phase layers, and the transmission coefficient of modulate layer

$m^{th}$ located at $z = z_m$ can be formulated as $t_m = a_m e^{i\varphi(x,y,z_m)}$, here we only consider the phase modulation case, i.e. $a_m = 1$. The optical field $f(x, y, z = z_m)$ modulated by $m^{th}$ diffractive layer at $z = z_m$ can be expressed as:

$$f(x, y, z = z_m) = t_m [f_{m-1}(x, y, z = z_{m-1}) * h(x, y, \Delta z_m)] \tag{3}$$

Where $\Delta z_m = z_m - z_{m-1}$ is the axial distance between two successive layers. After been modulated by all the layers, the intensity is presented at detection plane:

$$I(x, y) = |f(x, y, z = z_m) * h(x, y, \Delta z_m)|^2 \tag{4}$$

**2.2 Design of the random phase diffusers**

The random phase diffusers used for mimic complex media was modeled as thin phase mask, the transmittance is determined by using refractive index difference between air and diffusion material, the random height map $D(x, y)$ following the unit distribution. So, the complex transmittance can be expressed as

$$t_D = e^{i\frac{2\pi \Delta n}{\lambda} D(x,y)} \tag{5}$$

Where $\lambda$ is the wavelength of coherent illumination source, and random height map can be further expressed as

$$D(x, y) = W(x, y) * K(\sigma) \tag{6}$$

Where the $W(x, y)$ denotes the random height matrix, $K(\sigma)$ is the Gaussian smoothing kernel with standard deviation of $\sigma$, $*$ denotes 2D convolution operator. In the training process, $\mu = 25\lambda$ $\sigma_0 = 8\lambda$ and $\sigma = 4\lambda$ were set to randomly generate diffusers at each epoch.

**2.3 The design of network and training**

In the training process, the 2D space with a grid of $32\mu m$ was sampled, which is also the size of neuron, the size of input field of view (FOV) was set to be $12.6 \times 12.6 mm$, which corresponds to $240 \times 240$ pixels. Vortex beams used for simulated input field are generated by Python program, the phase value of diffusers is set from $0 \sim 2\pi$. The distance between layers is set to be $300\mu m$, During training phase, n phase diffusers with the same correlation length are randomly fed into diffractive model at each epoch. In each training iteration for single mode recognition, 10 VBs were duplicated n times and denotes VBs distorted by n random phase diffusers. So $B \times n$ distorted VBs are fed into network, and propagate through the diffractive layers. Eventually, different intensity block at the detection plane were presented, which were used for calculating loss function:

$$Loss = \frac{\sum_i^{Bn} \sum_{x,y} |I_{label}(x, y) - I_{pred}(x, y)|^2}{Bn} \tag{7}$$

In the optimization process, the gradient of the loss function with respect to all trainable network variables can be computed by using ADAM optimizer, and the values are then be utilized to update the network layers iteratively during each cycle of the training phase. So, we aim to jointly optimize the phase values to minimize the loss function:

$$\min loss(\phi_i), \quad 0 < \phi_i < 2\pi \tag{8}$$

In each iteration of the error backpropagation algorithm, a small batch of training data is used to feed into the diffractive neural network (D2NN). This batch consists of input samples along with their corresponding target outputs. By propagating the inputs forward through the network, the predicted outputs are obtained.

## 3. RESULTS AND DISCUSSION

Based on above model and parameter of network setting, we first simulated the free-space propagation of distorted VBs without the presence of diffractive layers, the VBs were fed into optical path and distorted by random diffusers, and the phase distribution of distorted VBs are presented in Fig.2. For illustrative purposes, we only present the phase distributions of vortex beams with topological charges of 5 and -3. One can clearly see in third column, the VBs propagation in freedom space and the wavefront are distorted by random diffusers, so that one can difficult to recognize the mode of VBs unless further digital calculations are performed on the computer.

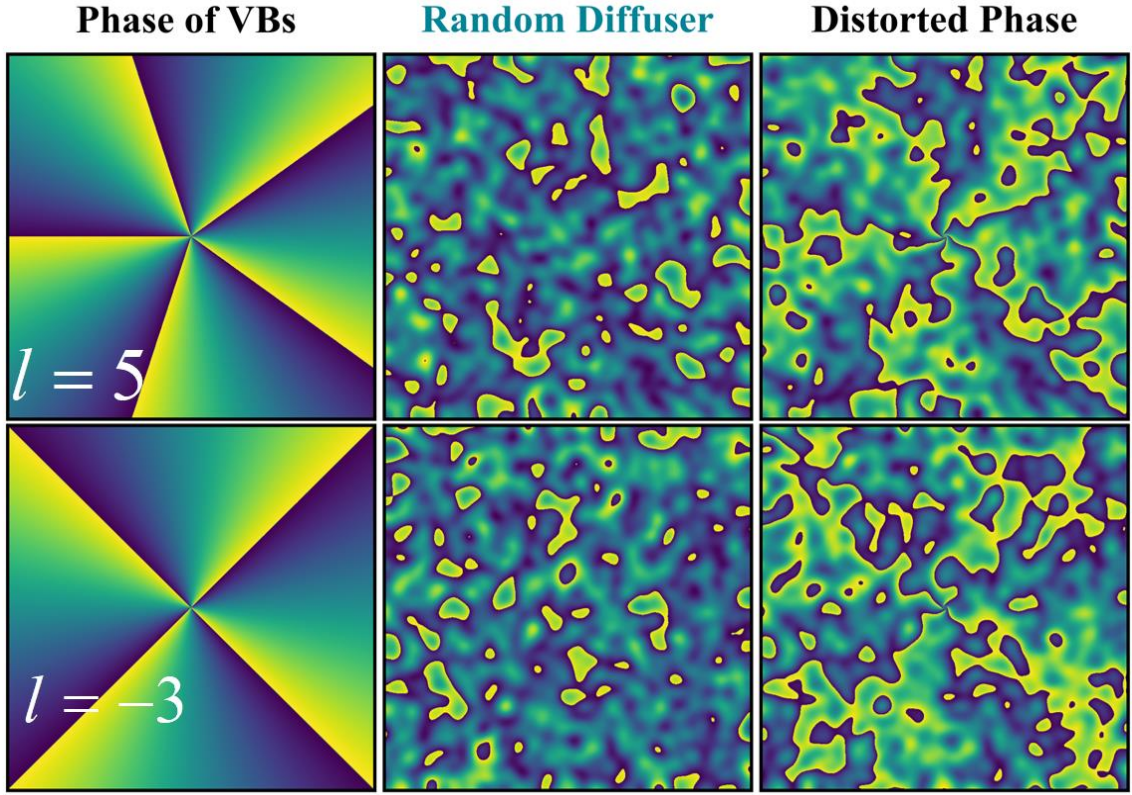

Figure.2 Propagation of VBs in freedom space and the phase distribution distorted by random phase diffuser along the optical path, the first column present the wavefront of VBs with ( $l = 5, -3$ ), the second column presents the random phase distribution with the same correlation length $L \sim 7\lambda$ that mimicking the glass-based diffusers, the third column presents the distorted phase distribution at the detection plane, due to the distortion of wavefront of VBs by scattering media, it is difficult to recognize the pattern unless further digital algorithms are involved.

To enhance the network's ability to generalize and optically discern the topological charge of vortex beams distorted by scattering media, multiple random diffusers with same correlation lengths ($L_{train} = L_{test} = 14\lambda$) are introduced into the network. These VBs propagate through the diffusers and are modulated by diffractive modulation layers to form the intensity distribution at the detection plane. Subsequently, the output results are utilized to compute the loss value. The Root Mean Square Error (RMSE), which quantifies the difference between predicted values and ground truth, was used to calculate the error gradients for updating the phase values of each neuron. To demonstrate the successful of our regime, we tested the random diffusers that were never seen during training stage. Under $\lambda = 620nm$ uniformly plane wave illumination, the input optical field propagate through diffusers and modulated by diffractive layers, and finally converge the diffractive energy to specific region on the detection plane, the physical plane of the network output was divided into discrete detection regions, each region of maximum optical intensity denotes the topological charge of the VBs. As shown in Fig 3, VBs with $l = 3$ can successfully be recognized by diffractive network, the region of intensity distribution represents the OAM mode. this indicates that our method for inference the mode of OAM can process the new random diffusers (never seen in training phase).

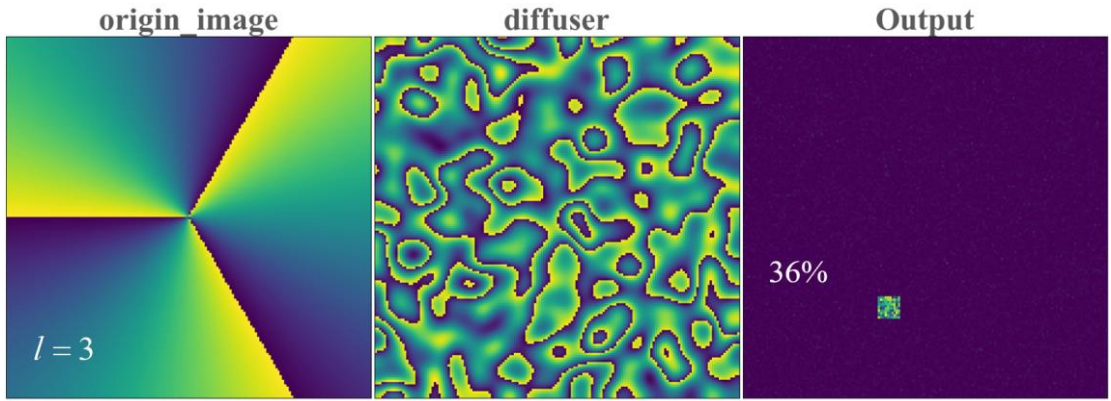

Figure3. Four-layer diffractive network trained to all- optically recognize the mode of VBs through random phase diffusers without any digital algorithm and computers. The first column denotes the phase distribution of original VBs that was not diatorted by diffusers, the midddle part represent random phase distribution of diffusers with correlation $L \sim 7\lambda$, the third column represent the intensity distribution at the detection plane, the position of the maximum intensity indicate the mode of VBs.

It is worth mentioning that although our network does not have an activation function, it still exhibits depth characteristics, which means that the increase the number of network layers will improve the inference capability of the optical network, specifically, it means increasing the percentage of light intensity within the target region. Throughout the optimization process, our goal is to fucus the intensity of the input light at a specific location by using optimization function. Despite our utilization of a limited subset of random phase diffusers, the successful recognition of VBs passing through new diffusers that were never seen in the training, furthermore, our model can avoids overfitting, exhibits a certain level of robustness.

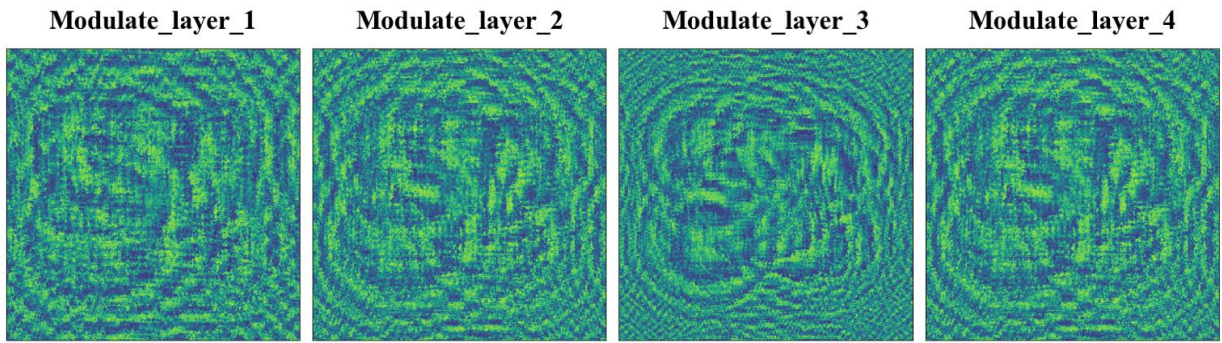

Figure 4 Phase patterns of the transmissive layers corresponding to the diffractive network that was trained using n=200 random diffusers at each epoch. The range of phase value in each neuron was set in the range of 0 to $2\pi$. The phase values of each neuron were adjusted during training process by using backpropagation algorithm.

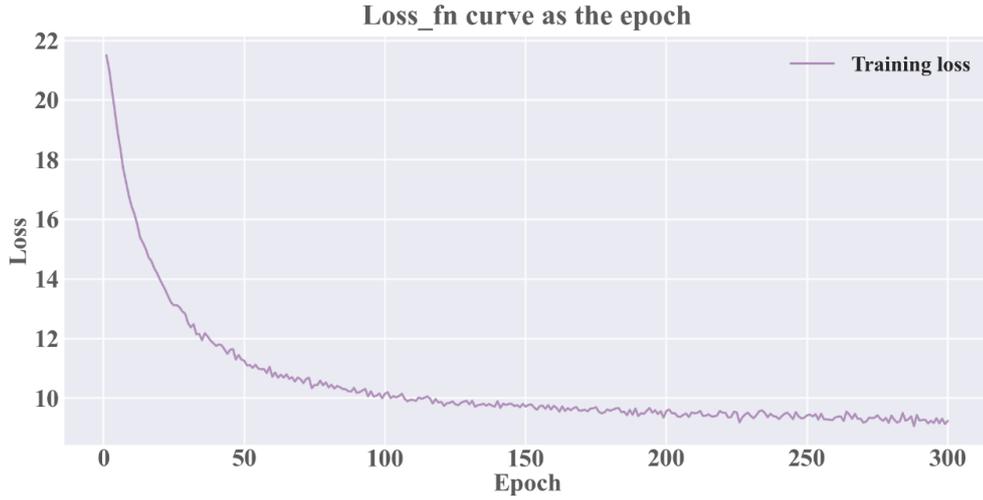

Figure 5 Loss value as the function of training epoch.

As demostrated in our numerical results, a diffractive network trained with random diffusers and VBs can all-optically recognize the OAM mode through new random diffusers, all the diffusers in the inference process are not involved in training dataset. Once training is completed, that is one time effort, the parameters of the network will be fixed, meaning the trained network can only perform specific task. Once deployed physically, the network can recognize mode of vortex beam passing through newly random diffusers at the speed of light. And zero energy consumption except for illumination source during the inference process. The physical implementation of the optical phase modulation layer can be achieved using spatial light modulators (SLMs) or diffractive optical elements. The choice of modulation unit size should ensure that the network adequately satisfies optical interconnections. Fig.4 shows the phase profiles of the trained diffractive layers. In the training phase, $200 \times 10$ random distorted VBs were fed into network at each epoch for extract pattern feature from the distorted vortex beam. Each neuron was ajustable parameter, and the phase value was set was set in the range of 0 to $2\pi$. Fig.5 shows the loss value as the function of the epoch, the decreasing loss function with increasing epochs indicates that the model gradually learns better parameters during the training process, resulting in predictions that are closer to the actual label (groundtruth), this reduction in loss typically reflects an improved fit of the model to the training data, meaning the inference of the model become more accurate over epoch.

## 4. CONCULUSION

In this study, we proposed an all-optical diffractive network-based approach for inferring orbital angular momentum (OAM) modes in vortex beams (VBs), enabling real-time recognition of distorted VBs without the need for digital reconstruction algorithms or computational processing. Our regime demonstrates that by introducing random phase diffusion layers to mimic scattering media, the trained model can achieve robust pattern recognition of OAM modes in an all-optical manner. While our investigation focuses on thin phase masks as proof of concept, we anticipate that our methodology can be extended to recognize OAM modes passing through volumetric diffusers, thereby addressing more complex and dynamic scattering scenarios. Ultimately, the insights and methods presented in our study offer valuable guidance for advancing vortex light communication technologies.